\documentclass[a4paper,11pt]{article}

\pdfoutput=1 

\usepackage{jinstpub} 

\usepackage{lineno}
\usepackage{multirow}
\usepackage[utf8]{inputenc}

\title{\boldmath Detection efficiency measurement and operational tests of the X-Arapuca for the first module of DUNE Far Detector}

\author[a,1]{C. Palomares,\note{Corresponding author.}}

\affiliation[a]{CIEMAT, Centro de Investigaciones Energ\'eticas, Medioambientales y Tecnol\'ogicas, \\ Av. Complutense 40, E-28040 Madrid, Spain}

\emailAdd{mc.palomares@ciemat.es}

\abstract{The Deep Underground Neutrino Experiment (DUNE) is a dual-site experiment for long-baseline neutrino oscillation studies, able to resolve the neutrino mass hierarchy and measure $\delta_{CP}$. 
DUNE will also have sensitivity to supernova neutrinos and to processes beyond the Standard Model, such as nucleon decay searches. 
The Far Detector (FD) will consist of four liquid argon TPC (17.5 kt total mass) with systems for the detection of charge and scintillation light produced by an ionization event. 
The charge detection system permits both calorimetry and position determination. 
In addition, the photon-detection system (PDS) enhances the detector capabilities for all DUNE physics drivers. The PDS of the first FD module consists of light collector modules placed in the inactive space between the innermost wire planes of the TPC anode. The light collectors, the so-called X-ARAPUCAS, are functionally a light trap that captures wavelength-shifted photons inside boxes with highly reflective internal surfaces where they are guided to Silicon Photo-multipliers (SiPM) by wavelength-shifting (WLS) bars. Functionality and operational tests of the X-ARAPUCAS to be installed in ProtoDUNE-SP phase II (FD DUNE prototype at the scale 1:20), as well as the measurement of their absolute detection efficiency is reported in this publication.}


\collaboration[c]{on behalf of DUNE collaboration}

\proceeding{N$^{\text{th}}$ LIGHT DETECTION IN NOBLE ELEMENTS (LIDINE 2022)\\
  21-23 SEPTEMBER, 2022 \\
  WARSAW (POLAND)}

\begin{document}
\maketitle
\flushbottom

\section{DUNE: Long-Baseline Neutrino Oscillation Experiment}
\label{sec:dune}

The Deep Underground Neutrino Experiment (DUNE)~\cite{dune_I} is a next-generation, long-baseline neutrino oscillation experiment which will carry out a detailed study of neutrino mixing utilizing high-intensity $\nu_{\mu}$ and $\bar{\nu}_{\mu}$ beams measured over a long baseline. DUNE is designed to make significant contributions to the completion of the standard three-flavor picture by measuring 
the parameters governing $\nu_{1}-\nu_{3}$ and $\nu_{2}-\nu_{3}$ mixing and the neutrino mass ordering in a single experiment. Paramount among these is the search for charge-parity symmetry violation in neutrino oscillations.
Other primary science goals are search for proton decay and detect and measure the $\nu_{e}$ flux from a core-collapse supernova within our galaxy.

\begin{figure}[htbp]
\centering 
\includegraphics[width=0.8\textwidth,trim=0 0 0 100,clip]{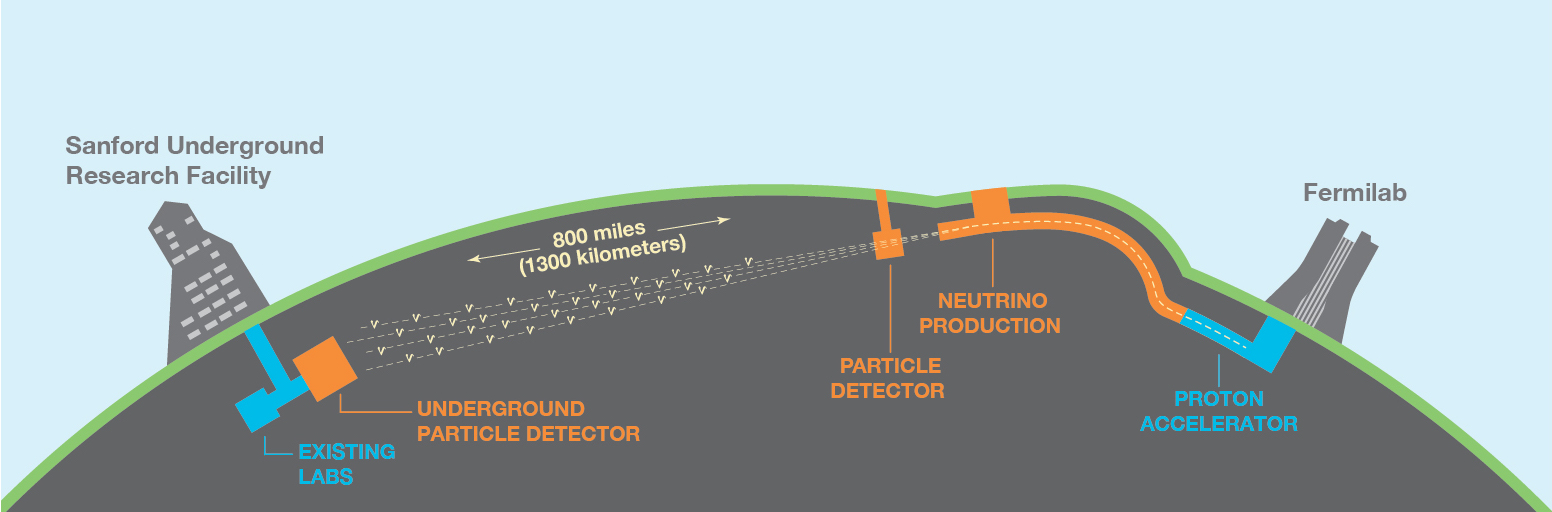}
\caption{\label{fig:dune} DUNE overview as dual-site experiment.}
\end{figure}

DUNE will consist of a far detector to be located about 1.5 km underground at the Sanford Underground Research Facility (SURF) in South Dakota, USA, at a distance of 1300 km from Fermilab, where a near detector will be located (figure~\ref{fig:dune}).
The DUNE experiment will observe neutrinos from a high-power $\nu_{\mu}$ and $\bar{\nu}_{\mu}$ beam peaked at $\sim$2.5 GeV but with a broad range of neutrino energies.

\subsection{First Module of DUNE Far Detector}

\begin{figure}[htbp]
\centering 
\includegraphics[width=0.45\textwidth,trim=5 0 5 5,clip]{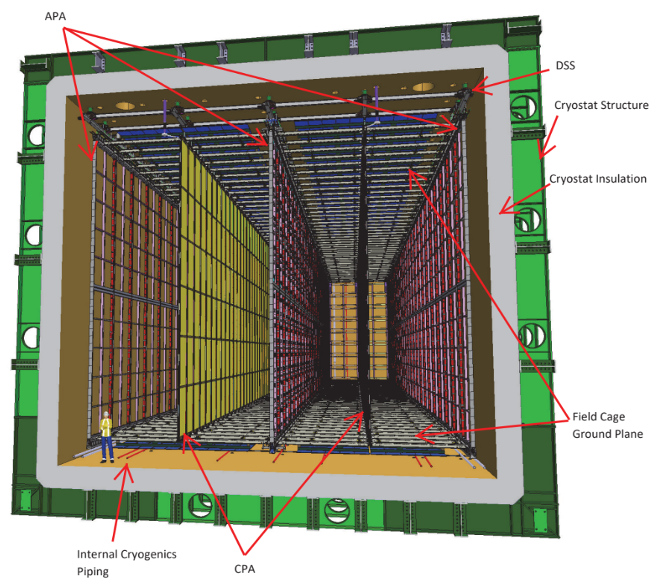}
\includegraphics[width=0.45\textwidth,trim=5 0 5 5,clip]{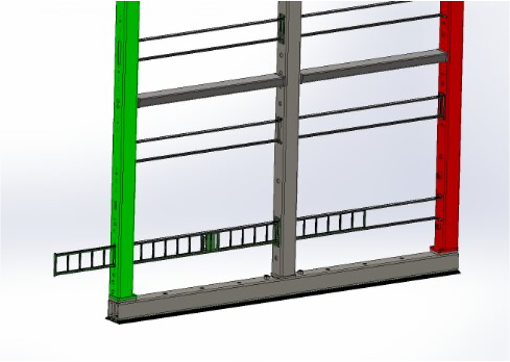}
\caption{\label{fig:fd_drift} Left: A 10 kt DUNE far detector module, showing the alternating 58.2 m long, 12.0 m high anode and cathode planes (APA and CPA), as well as the field cage that surrounds the
drift regions between the anode and cathode plane. Right: X-ARAPUCA module mounted inside an anode plane.}
\end{figure}

The DUNE far detector (FD) will consist of four liquid argon time-projection chambers (LArTPC) with a total mass of nearly 70~kt (fiducial mass of at least 40~kt). This liquid argon (LAr) technology will make possible the reconstruction of neutrino interactions with image-like precision.
The design of the four identically sized modules is sufficiently flexible for staging construction and evolving the LArTPC technology.
The first FD Module will use the single-phase (SP) technology~\cite{dune_fd}, in which ionization charges drift horizontally in the LAr under the influence of an electric field towards a vertical anode, where they are read out. Four 3.5 m drift volumes are created between five alternating anode and cathode walls, each wall having dimensions of 58 m $\times$ 12 m, and installed inside a cryostat, shown in figure~\ref{fig:fd_drift}-left.

\subsection{Photon-detection System}

\begin{figure}[htbp]
\centering 
\includegraphics[width=0.12\textwidth,trim=5 0 5 5,clip]{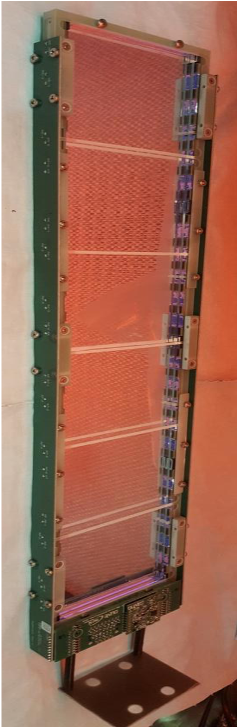}
\hspace{2cm}\includegraphics[width=0.25\textwidth,trim=0 0 0 0,clip]{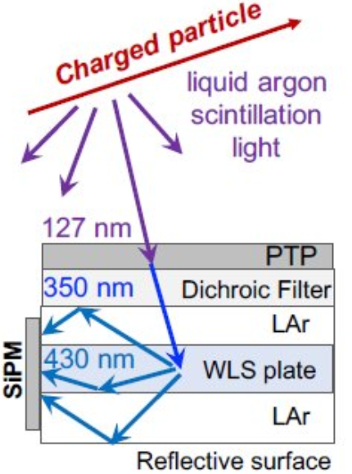}
\caption{\label{fig:pds} Left: Image of one of the X-ARAPUCAs tested for ProtoDUNE. Right: Schematic of the X-ARAPUCA working principle.}
\end{figure}

The photon detection system (PDS) must measure the VUV scintillation light produced by ionizing tracks in the TPC within the geometrical constraints of the anode plane structure. 
The detection of the scintillation light will enhance DUNE detector capabilities.
The light collector modules are placed in the inactive space between the innermost wire planes of the anode planes, as it is indicated in figure~\ref{fig:fd_drift}-right.  The X-ARAPUCA is an improvement of the original concept of photon trapping inside a highly reflective box while using a wavelength shifting (WLS) bar to increase the probability of collecting trapped photons onto a SiPM array. They have an outer layer of p-Terphenyl (pTP) that converts the incident 127 nm scintillation photons into 350 nm wavelength. Then the light reaches the WLS bar and it is shifted to the visible range (430 nm) above the dichroic filter cut-off of 400 nm, as illustrated in the schematic of figure~\ref{fig:pds}-right. The WLS re-emitted light can be trapped by total internal reflection or escape and be reflected on the dichroic filter, reaching at the end the Silicon Photo-Multiplier (SiPM) photo-sensors.
The performance required for the PDS to achieve 99\% for tagging nucleon decay events is a light yield of 0.5 PE/MeV at the furthest point (near the cathode plane), while the requirement to enable a calorimetric energy measurement with the PDS for low-energy events like Supernova burst is 20 PE/MeV averaged over the active volume, which corresponds to a collection efficiency of the single cell of 1.3\% and 2.6\%, respectively.

\paragraph{ProtoDUNE-SP Phase II}
In order to validate the DUNE technology, prototypes are being designed, built and tested at CERN, the so-called ProtoDUNEs~\cite{proto_sp}. 
ProtoDUNE-SP phase II will test the final design of the first FD module at scale 1:1 in 2023. In particular, the X-ARAPUCAs will be tested for the first time.
For their installation in ProtoDUNE phase II and in the first FD module, the X-ARAPUCA is composed by six coated filters, a WLS bar and 48 SiPMs, as can be seen in the picture featured in figure~\ref{fig:pds}-left.
Different configurations of these X-ARAPUCAs depending on the model of SiPMs or WLS bar chosen will be tested in ProtoDUNE. Two different models of SiPMs, the first ones manufactured by Fondazione Bruno Kessler (FBK) and the second ones by Hamamatsu Photonics K.K (HPK), will be employed. In particular, they are: FBK Triple-Trench (TT), which pixel size is about 50 $\mu$m and HPK 75 $\mu$m High Quenching Resistance (HQR) both with a total effective area of 36 mm$^2$ and specifically designed for being used at cryogenics temperatures. ProtoDUNE phase II also makes use of two different models for the WLS bar, the EJ-286PS manufactured by Eljen Technology and the bar designed by Glass to Power (G2P) in collaboration with INFN.

\section{Measurement of X-Arapuca Efficiency}

Quantifying the absolute efficiency of the X-ARAPUCA to VUV scintillation light is fundamental for DUNE, as the parameter is needed to fully characterize the photon-detection system. For that purpose, we submerged an X-ARAPUCA in LAr together with a low-activity electro-deposited $^{241}{\rm Am}$ alpha source to expose the device to scintillation light of 127 nm.

Two dedicated setups have been developed, at CIEMAT (Madrid-Spain) and at Milano-Bicocca University (Italy). Both of them liquefy the gas argon (GAr) in the inner most vessel at the expenses of the evaporation of liquid nitrogen or argon of an external bath.

To measure the absolute efficiency of the X-ARAPUCA, the number of photons arriving to its surface is needed. Two different methods have been considered: In the so-called {\it Method A}, the amount of light collected by the X-ARAPUCA is compared with the light detected by a calibrated photo-sensor. In the {\it Method B}, the light in the X-ARAPUCA is estimated from the $\alpha$-source energy and the known number of scintillation photons per MeV produced in LAr once the solid angle sustained by the source is determined.   

\subsection{CIEMAT setup}

\begin{figure}[htbp]
\centering 
\includegraphics[width=0.25\textwidth,trim=5 30 5 5,clip]{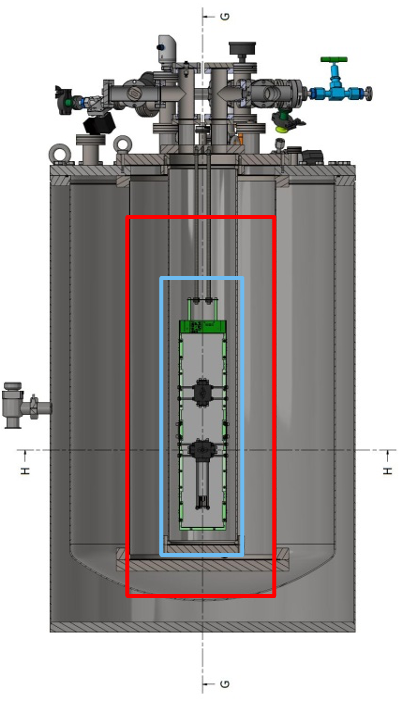}
\includegraphics[width=0.45\textwidth,trim=0 0 0 0,clip]{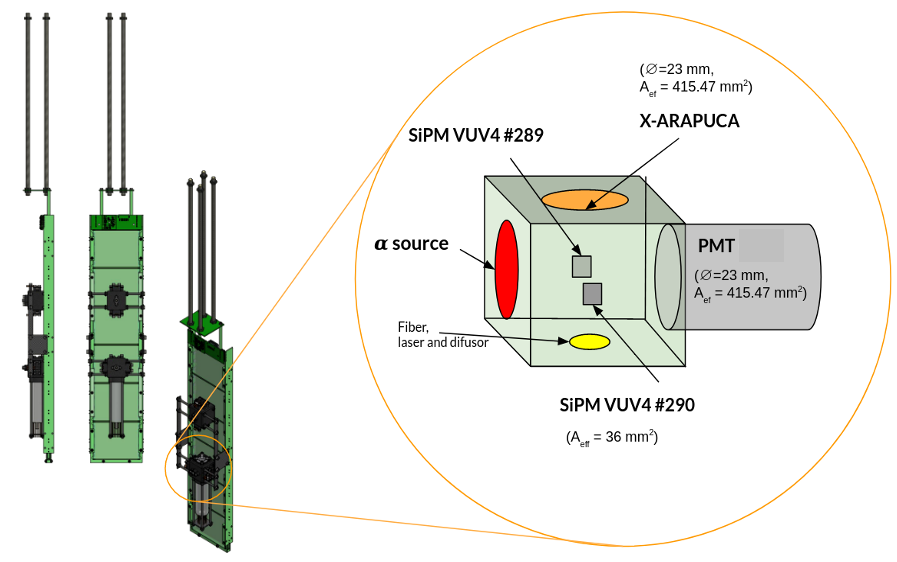}
\caption{\label{fig:setup} CIEMAT setup scheme used for measuring X-ARAPUCA efficiency. Left: Concentric vessels. The vessel containing LAr is marked in blue, while the red one is filled with LN2. Right: Small black box holding the $\alpha$-source together with different photo-sensors.}
\end{figure}

The CIEMAT neutrino group made use of a 300 L cryogenic vessel with different concentric volumes, which schematic is shown in figure~\ref{fig:setup}. There are two concentric internal volumes, a larger one (100 L), where the liquid nitrogen (LN2) is introduced and a smaller one, 18 L, where the X-ARAPUCA is located and is filled with gas argon (GAr). In this 18 L container, GAr 99.9999\% is liquefied with the use of LN2. This is achieved by controlling the pressure at 2.7 bar. 
To avoid outgassing, successive vacuum cycles are performed before introducing the optical and electrical components.
The X-ARAPUCA is submerged in the inner vessel together with the two reference calibrated VUV4 SiPMs (model S13370-6075CN) and a VUV-sensitive Photo-Multiplier Tube (PMT) to monitor the decay time of the scintillation slow component and indeed the LAr purity. The VUV4 SiPMs are designed to have a high sensitivity for VUV light
and they are prepared to carry out a stable performance at cryogenic temperature. 
The $^{241}$Am source is held in a 2 cm size opaque box, as shown in figure~\ref{fig:setup}, where the $\alpha$ particles deposit their energy, and the emitted photons will reach the X-ARAPUCA through a hole ($\oslash = 23$ mm) that faces it. We ensure that no other photons are being detected by covering the rest of the X-ARAPUCA with a black sheet. On the other faces of the box, we place the two VUV4 SiPMs and the PMT. 

\subsection{Milano Bicocca University setup}
\begin{figure}[htbp]
  \centering 
  \includegraphics[width=0.12\textwidth,trim=0 0 0 0,clip]{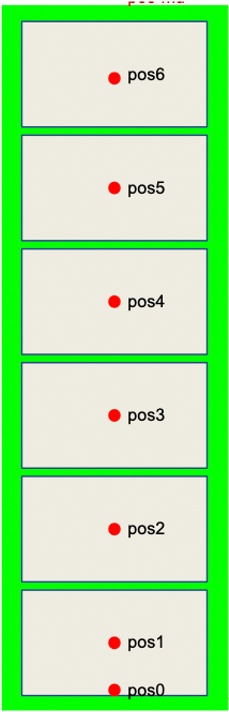}
\includegraphics[width=0.4\textwidth,trim=0 0 0 0,clip]{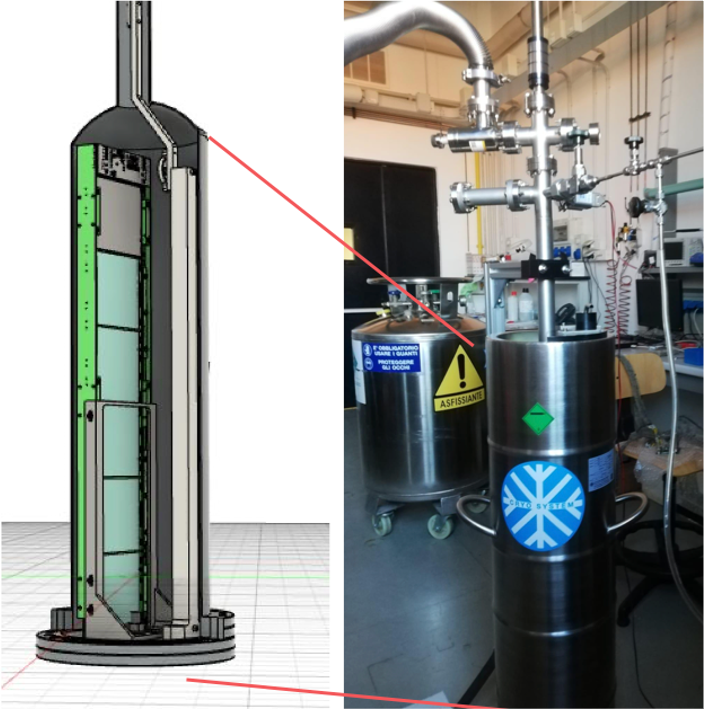}
\includegraphics[width=0.46\textwidth,trim=0 0 0 0,clip]{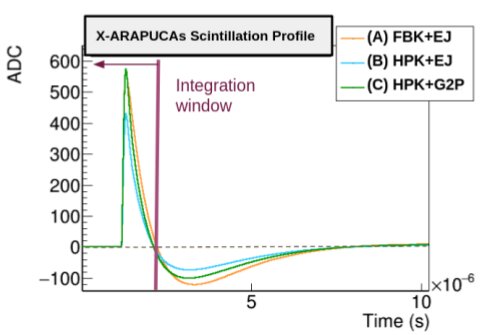}
\caption{\label{fig:setup2} Left: The seven height positions at fixed distance from the X-ARAPUCA used for its characterization at MiB. Pos0 corresponds to the most unfavorable case. Centre: 3D drawing of the chamber with the X-ARAPUCA and the source sliding rail, and an image of the cryogenic vessel used at MiB. Right: Typical waveform obtained from the X-ARAPUCA response to the scintillation light produced by an $\alpha$ particle.}
\end{figure}

The setup developed by the Milano Bicocca University (MiB)~\cite{mib_jinst} is shown in figure~\ref{fig:setup2}-centre. A cryogenic stainless steel chamber sizing 250 mm diameter by 310 mm height, for a total volume of $\sim$5 liters. All the sensors and safety devices are stainless steel, whole-metallic, to minimize the outgassing in the chamber. Prior to the gas-Ar (GAr) liquefaction, the chamber is first pumped down to $\sim 10^{-3}$ mbar, then the GAr (6.0 grade) inlet is opened, while the chamber stays immersed in an external LAr bath. The pressure gradient generated by the liquefaction process, makes the GAr continuously flow from the bottle to the chamber where it is liquefied. The pressure in the chamber is regulated at about 1.4 bar, allowing to fill it in about 3 hours.
The $\alpha$ source was fixed on the rail at a distance of $5.5\pm 0.2$ cm from the dichroic filters and was moved to six different positions in Z to scan the X-ARAPUCA as shown in figure~\ref{fig:setup2} left. 

\subsection{Method A}

This method is used by CIEMAT group. The efficiency of the X-ARAPUCA, $\epsilon_{A} (\rm Arapuca)$,  is obtained from the known efficiency of the reference SiPMs, $\epsilon(Ref. SiPM)$, as is shown in equation~\ref{eq:mtha}.

\begin{equation}
\label{eq:mtha}
\epsilon_{A} (\rm Arapuca) = \frac{\#PE_{mm^{2}}(Arapucas)}{\#PE_{mm^{2}}(Ref. SiPM)}\cdot\epsilon(Ref. SiPM)\cdot f_{corr}   
\end{equation}

\noindent where, $\#PE_{mm^{2}}$ is the number of photo-electrons (PE) per area in mm$^{2}$ and $f_{corr}$ is a correction factor that accounts for: (1) the different cross-talk of the X-ARAPUCA SiPMs and reference VUV4 SiPMs, (2) the fraction of the collected light by the X-ARAPUCA that is not included in the waveform integration due to an undershooting in the pulse (figure~\ref{fig:setup2}-right), and (3) the different solid angle for the X-ARAPUCA and the reference SiPMs due to the different surface areas.

\subsection{Method B}

This method has been chosen by MiB group and it used as cross-check by CIEMAT group. The scintillation light at the X-ARAPUCA surface is determined with equation~\ref{eq:mthb:2}, while the X-ARAPUCA efficiency is calculated using equation~\ref{eq:mthb:1}.

  \begin{subequations}\label{eq:mthb}
\begin{equation}
\label{eq:mthb:1}
\epsilon_{B} (\rm Arapuca)  = \frac{\#PE(Arapucas)}{\#Ph}\cdot f'_{corr}
\end{equation}

\begin{equation}
\label{eq:mthb:2}
\#{\rm Ph}  = LY_{LAr}\cdot E_{\alpha} = 35000 {\rm ph}/{\rm MeV}\cdot 5.48 {\rm MeV}\cdot\Omega
\end{equation}
  \end{subequations}

 \noindent the light yield has been obtained from~\cite{doke} assuming the $\alpha$ quenching factor of 0.7~\cite{mei}. $f'_{corr}$ in this case must correct not only the SiPMs cross-talk effect and the fraction of light not included in the waveform integration but also the light production quenching by impurities. 

  \subsection{Results}
  The absolute efficiency has been measured for different SiPM and WLS bar configurations. The results obtained in the two labs and through the two proposed methods are shown in Table~\ref{tab:2}. In {\it Method A}, the main error source is the uncertainty in the reference SiPM PDE; while in {\it Method B}, it is the source positioning.
  
  MiB group has observed a non-uniformity in the efficiency along the X-ARAPUCA, more pronounced for the configurations including FBK SiPMs, the quoted results correspond to the average of the seven measurements done along the device. By contrast, the CIEMAT measurements correspond to the center of the X-ARAPUCA. That would explain the lower value of the efficiencies obtained by MiB group. 

  In the case of CIEMAT measurement, the efficiencies obtained with {\it Method A} are a $\sim37\%$ higher than with {\it Method B}; this could be due to an overestimation of the reference SiPM PDE. Theses SiPMs were calibrated at room temperature by the manufacturer; however, several measurements, including some done at CIEMAT, shown a PDE at cryogenic temperature a $\sim50\%$ lower. The value used in this analysis has been taken from a recent publication~\cite{vuv4_sipm}.
  
  \begin{table}[htbp]
\centering
\caption{\label{tab:2} Results for the absolute efficiency ($\epsilon$) of the different X-ARAPUCAs' configurations for a SiPM bias voltage corresponding to a PDE of 45\%.}
\smallskip
\begin{tabular}{|c|c|c|c|c|c|}
\hline
LAB & PDE$_{XA-SiPM} = 45$\% & FBK + EJ & FBK+G2P & HPK+EJ & HPK+G2P \\
\hline
\multirow{2}{*}{CIEMAT} &  $\epsilon_{A}$ (\%) & $2.12\pm 0.15$ & & $2.36\pm 0.17$ & $3.15\pm 0.23$ \\
&  $\epsilon_{B}$ (\%) & $1.56\pm 0.12$ & & $1.72\pm 0.14$ & $2.28\pm 0.19$ \\
\hline
MiB & $\epsilon_{B}$ (\%) & $1.29\pm 0.07$ & $1.49\pm 0.10$ & & $1.82\pm 0.08$ \\
\hline
\end{tabular}
\end{table}

  \section{Conclusions}
  In this publication, we report the measurement of the absolute detection efficiency of the X-ARAPUCAs to be installed in ProtoDUNE-SP phase II. For all the configurations tested, the efficiency, for a SiPM bias voltage corresponding to 45\% PDE, lays in between 1.3\% and 3\%. Achieving the performance required for the photon-detection system.


\begin{thebibliography}{99}

\bibitem{dune_I}
DUNE Collab., B. Abi et al, \emph{Deep Underground Neutrino Experiment (DUNE), Far Detector Technical Design Report, Volume I Introduction to DUNE}, \emph{JINST} {\bf 15} 08 (2020) T08008.

\bibitem{dune_fd}
DUNE Collab., B. Abi et al, \emph{Deep Underground Neutrino Experiment (DUNE), Far Detector Technical Design Report, Volume IV: Far Detector Single-phase Technology} \emph{JINST} {\bf 15} 08 (2020) T08010.

\bibitem{proto_sp}
  DUNE Collab., B. Abi et al, \emph{First results on ProtoDUNE-SP liquid argon time projection chamber performance from a beam test at the CERN Neutrino Platform}  \emph{JINST} {\bf 15} 12 (2020) P12004.

\bibitem{mib_jinst}
C. Brizzolari et al., \emph{Enhancement of the X-Arapuca photon detection device for the DUNE experiment} \emph{JINST} {\bf 16} 09 (2021) P09027.
\bibitem{doke}
T. Doke,  \emph{Fundamental Properties of Liquid Argon, Krypton and Xenon as Radiation Detector Media.} \emph{Experimental Techniques in High-Energy Nuclear and Particle Physics} (1991) pp. 537-577.
\bibitem{mei}
  D.-M. Mei, Z.-B. Yin, L. Stonehill and A. Hime, \emph{A model of nuclear recoil scintillation efficiency in noble liquids} \emph{Astroparticle Physics} {\bf 30} (2008) 12–17.
\bibitem{vuv4_sipm}
  T. Pershing, J. Xu, E. Bernard, J. Kingston, E. Mizrachi, J. Brodsky, A. Razeto, P. Kachru, A. Bernstein, E. Pantic and I. Jovanovic, \emph{Performance of Hamamatsu VUV4 SiPMs for detecting liquid argon scintillation} \emph{JINST} {\bf 17} 04 (2022) P04017.

  


\end{thebibliography}
\end{document}